%% file: main.tex
\documentclass[10pt]{article}
\usepackage{setspace}
\usepackage[margin=1.1in]{geometry}

\usepackage[font=footnotesize]{caption}

\usepackage{amsgen}
\usepackage{amsfonts}
\usepackage{amssymb}
\usepackage{amsbsy}

\usepackage{graphicx}
\usepackage[breaklinks=true,colorlinks=true,linkcolor=blue,urlcolor=blue,citecolor=blue]{hyperref}
\usepackage[ruled,vlined]{algorithm2e}
\usepackage{algorithmic}
\usepackage{color}
\usepackage{multirow}

\usepackage{amsmath,amssymb}
\DeclareMathOperator{\Lagr}{\mathcal{L}}
\DeclareMathOperator{\E}{\mathbb{E}}
\DeclareMathOperator*{\argmin}{arg\,min} 
\DeclareMathOperator*{\argmax}{arg\,max} 
\usepackage{caption}
\usepackage{subcaption}

\newcommand{\norm}[1]{\left\lVert#1\right\rVert}

\title{Multitask 3D CBCT-to-CT Translation and Organs-at-Risk Segmentation Using Physics-Based Data Augmentation}
\author{Navdeep~Dahiya$^{1*}$,~Sadegh~R~Alam$^{2}$,~Pengpeng~Zhang$^{2}$,~Si-Yuan~Zhang$^{3}$,\\~Tianfang~Li$^2$,~Anthony~Yezzi$^{1}$,~and~Saad~Nadeem$^{2}$}
\date{}

\usepackage{fancyhdr}
\begin{document}
\maketitle

{\footnotesize
\noindent
$^1$Department of Electrical \& Computer Engineering, Georgia Institute of Technology, Atlanta, GA, USA.

\noindent
$^2$Department of Medical Physics, Memorial Sloan-Kettering Cancer Center, New York, NY, USA.

\noindent
$^3$Department of Radiation Oncology, Peking University Cancer Hospital, Beijing, China.

\noindent
$^*$ Work done as an intern at MSKCC.

\noindent
\textbf{Correspondence:} Saad Nadeem (nadeems@mskcc.org)
}\\

\begin{abstract}
\noindent
\textbf{Purpose:} In current clinical practice, noisy and artifact-ridden weekly cone-beam computed tomography (CBCT) images are only used for patient setup during radiotherapy. Treatment planning is done once at the beginning of the treatment using high-quality planning CT (pCT) images and manual contours for organs-at-risk (OARs) structures. If the quality of the weekly CBCT images can be improved while simultaneously segmenting OAR structures, this can provide critical information for adapting radiotherapy mid-treatment as well as for deriving biomarkers for treatment response.\\
\noindent
\textbf{Methods:} Using a novel physics-based data augmentation strategy, we 
synthesize a large dataset of perfectly/inherently registered planning CT and synthetic-CBCT pairs for locally advanced lung cancer patient cohort, which are then used in a multitask 3D deep learning framework to simultaneously segment and translate real weekly CBCT images to high-quality planning CT-like images.\\
\noindent
\textbf{Results:} We compared the synthetic CT and OAR segmentations generated by the model to real planning CT and manual OAR segmentations and showed promising results. The real week 1 (baseline) CBCT images which had an average MAE of 162.77 HU compared to pCT images are translated to synthetic CT images that exhibit a drastically improved average MAE of 29.31 HU and average structural similarity of 92\% with the pCT images. The average DICE scores of the 3D organs-at-risk segmentations are: lungs 0.96, heart 0.88, spinal cord 0.83 and esophagus 0.66. \\
\noindent
\textbf{Conclusions:} We demonstrate an approach to translate artifact-ridden CBCT images to high quality synthetic CT images while simultaneously generating good quality segmentation masks for different organs-at-risk. This approach could allow clinicians to adjust treatment plans using only the routine low-quality CBCT images, potentially improving patient outcomes. Our code, data, and pre-trained models will be made available via our physics-based data augmentation library, Physics-Arx, at \url{https://github.com/nadeemlab/Physics-ArX}.
\end{abstract}



\section{Introduction}
\input{introduction}

\section{Materials and Method}
\label{sec:materialsandmethods}
\input{materialsmethods}

\section{Results}
\label{sec:results}
\input{experimentalresults}

\section{Discussion}
\label{sec:discussion}
\input{discussion}

\section{Conclusion}
\label{sec:conclusion}
\input{conclusions}

\section*{Acknowledgments}
This project was supported by MSK Cancer Center Support Grant/Core Grant (P30 CA008748). This work was funded in part by National Institutes of Health (NIH) grant R01 HL143350.

\section*{Financial Disclosures}
The authors have no conflicts to disclose.

\section*{Code and Data Availability}
The code and the trained models will be made available via our physics-based data augmentation library, Physics-Arx, at \url{https://github.com/nadeemlab/Physics-ArX}. Physics-augmented AAPM data used for training our model will be made available via \url{https://zenodo.org/record/5002882#.YM_ukXVKiZQ}. Internal processed/de-identified data can be made available via a reasonable request.

\bibliographystyle{medphy}

\clearpage
\section*{Appendix}
\input{appendix}
\end{document}

%% file: introduction.tex
In image-guided radiotherapy~\cite{Boda-Heggemann2011,Elsayad2016} for lung cancer treatment, weekly cone-beam computed tomography (CBCT) images are primarily used for patient setup. At the start of treatment, high quality planning computed tomography (pCT) images are acquired for treatment planning purposes. Tumors and major organs-at-risk (OARs) are manually delineated by a trained radiation oncologist using the pCT images. During the treatment phase, weekly/daily low-quality CBCT images are acquired for patient setup and qualitative visual assessment of tumor and critical OARs. A variety of scattering and noise artifacts~\cite{doi:10.1259/dmfr/30642039} render CBCT images unsuitable for quantitative analysis (e.g., it is much harder to manually delineate OARs in CBCT images due to low soft tissue contrast). 

Several attempts have been made to quantify CBCT images in radiotherapy (RT) using both model-based methods~\cite{Zhu2009,Xu2015,Wu2014} as well as the more recent deep learning based methods~\cite{8331163,xu2018deep}. CBCT imaging suffers from several types of artifacts and these methods focus on the correction of only a particular type of artifacts such as either beam hardening~\cite{Li2016,Zhao_2011}, scattering~\cite{Xu2015}, metal artifacts~\cite{Wu2014}, or cupping~\cite{Xie2016}. Instead of trying to fix particular noise artifacts in CBCT images, a more recent line of research using deep learning methods attempts to directly generate higher-quality synthetic CT (sCT) from CBCT images. One particular approach is to use cycle-consistent generative adversarial networks~\cite{CycleGAN2017} (CycleGAN) to generate sCT from CBCT images~\cite{Liang_2019,Kurz_2019}. CycleGANs and other unsupervised image-to-image translation methods using unpaired CT and CBCT images can easily produce randomized outputs or in other words, hallucinate anatomy \cite{cohen2018distribution}. 

A supervised learning based method, sCTU-net~\cite{https://doi.org/10.1002/mp.13978}, used a 2D Unet~\cite{ronneberger2015unet} architecture to translate CBCT to sCT images. In addition to 2D slices of weekly CBCT images, it also feeds the corresponding pCT slice to the network and uses a combination of mean absolute error and structural similarity index loss to train the network. It utilizes hand-crafted loss function which makes it difficult to enhance this method for a multitask setting of simultaneous CBCT-to-CT translation and OARs segmentation. This is due to the fact that addition of a loss term for segmentation may cause destructive gradient interference during back-propagation. Another recent approach~\cite{Cycle-DeblurTien2021} combined the architecture of a GAN designed to deblur images~\cite{Deblur-GAN} with the CycleGAN architecture~\cite{CycleGAN2017} to translate chest CBCT images to synthetic CT images. The authors showed better results compared to CycleGAN alone~\cite{CBCT-CT-CycleGAN} as well as to a residual encoder-decoder convolution neural network (RED-CNN)~\cite{RED-CNN}, designed to reduce noise in low-dose CT images. Another method designed for segmenting prostate CBCT images~\cite{Prostate-Seg-10.1117/12.2580791} trained a CycleGAN model using aligned CBCT and MRI pairs to translate prostate CBCT images to synthetic MRI (sMRI) images. Afterwards it used two different Unet models to extract features from the CBCT and sMRI images respectively, combined the features using attention gates and finally predicted multi-organ prostate segmentations using a CNN network. A recent paper~\cite{Generalizability-Liang_2020} studied generalizability issues in deep learning based models in applying a model trained on a particular dataset to another different dataset. The authors investigated how a model trained for one machine and one anatomical site works on other machines and other anatomical sites for the task of CBCT to CT translation. The paper also explored solutions for the generalizability issues using the transfer learning approach.

In this paper, we use a supervised image-to-image translation method based on conditional generative adversarial networks~\cite{mirza2014conditional} (cGANs), to translate CBCT to sCT images while also performing OAR segmentation driven by a novel physics-based artifact/noise-induction~\cite{Sadegh-psCBCT} data augmentation pipeline. A particular single-task 2D implementation of cGANs called \textit{pix2pix}~\cite{pix2pix2017} is a generic image translation approach which obviates the need for hand-engineered loss functions and unlike CycleGANs does not produce randomized outputs. The physics-based data augmentation technique creates multiple perfectly paired/registered pseudo-CBCTs (psCBCTs) corresponding to a single planning CT. Combined with geometric data augmentation techniques, this non-deep learning strategy allows us to generate several perfectly registered pCT/CBCT/OARs pairs from a single pCT. This makes it possible for us to train a multi-task 3D cGAN based model inspired by \textit{pix2pix} for the joint task of CBCT to sCT translation and OARs segmentation without the risk of over-fitting. To further improve our results, we used our psCBCT generation technique on an open source dataset used in American Association of Physicists in Medicine (AAPM) thoracic auto-segmentation grand challenge~\cite{https://doi.org/10.1002/mp.13141}. We converted the AAPM planning CT and OAR segmentation dataset consisting of 60 cases to an augmented psCBCT dataset using artifacts extracted from one of our internal CBCT images and used it to further augment our training data that resulted in better segmentation and translation performance.

%% file: materialsmethods.tex
\subsection{Input Data}
\begin{figure}[htb!]
    \centering
    \includegraphics[width=\linewidth]{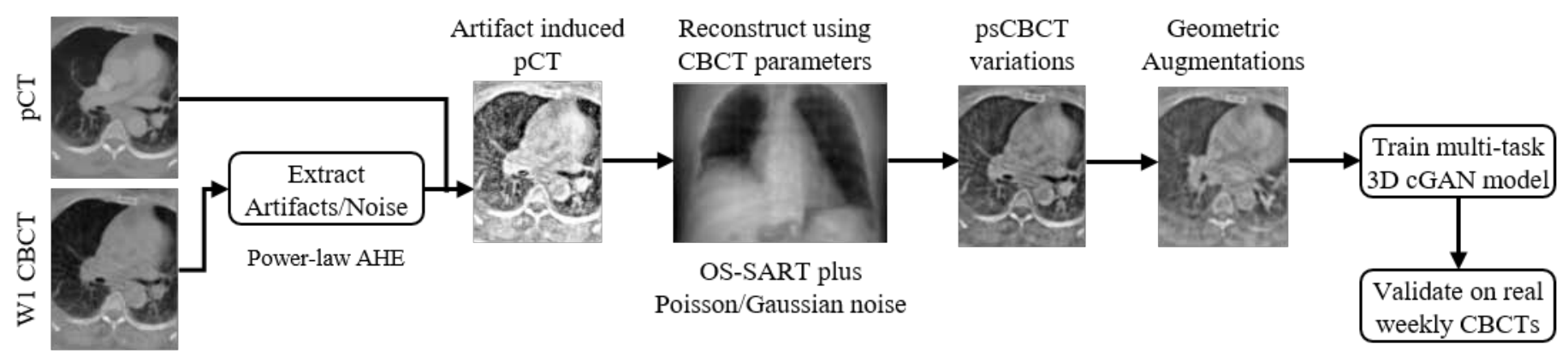}%
    \caption{Entire workflow of generating different variations of perfectly paired pseudo CBCT variations from single deformable registered planning CT and week1 CBCT pair. The generated data is used to train a multitask 3D model to generate synthetic CT and OAR segmentations. The trained model is tested on real weekly CBCT data.}
    \label{fig:psCBCTPipeline}
\end{figure}

We use a novel physics-based artifact induction technique to generate a large dataset of perfectly-paired/registered pseudo CBCTs (psCBCTs) and corresponding planning CTs (pCTs)/organs-at-risk (OARs) segmentations. A variety of CBCT artifacts are extracted from real week 1 (baseline) CBCTs and mapped to corresponding pCTs to generate psCBCTs. The generated psCBCTs have similar artifacts distribution as the clinical CBCT images and include all the physics-based aspects of diagnostic imaging i.e. scatter, noise, beam hardening, and motion. This data augmentation technique creates multiple paired/registered psCBCTs corresponding to a single planning CT. Figure~\ref{fig:psCBCTPipeline} shows our entire workflow for generating multiple perfectly paired psCBCT variations from a single pair of deformably registered week1 CBCT and pCT pair.

\subsubsection{Image Dataset}
We used the data from a study which included 95 locally advanced non-small cell lung cancer patients treated via intensity-modulated RT and concurrent chemotherapy. All patients had high-quality planning CT and 5/6 weekly CBCTs; all CTs were 3D scans acquired under free-breathing conditions. The pCT and wCBCT resolutions were $1.17\times 1.17\times 3.0\, mm^3$ and $0.98\times 0.98\times 3.0\, mm^3$ respectively. OARs including, esophagus, spinal cord, heart and lungs were delineated by an experienced radiation oncologist according to anatomical atlases of organs-at-risk~\cite{Kong2011}. This data formed the basis for generating synthetic artifact-induced psCBCTs which in turn was used to train our multitask cGAN model.

\subsubsection{Pseudo-CBCT Dataset}
\begin{figure}[htb!]
    \centering
    \includegraphics[width=\linewidth]{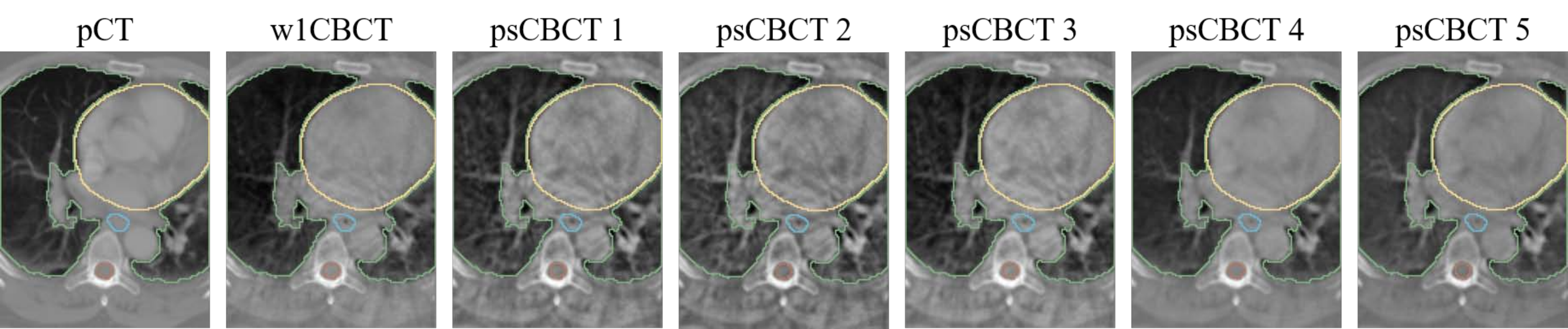}%
    \caption{Example result of the physics-based psCBCT generation process. Left two columns show several slices of deformably registered pCT and real week 1 CBCT images. The next 5 columns show the corresponding slices of five different variations of psCBCT images generated by transferring artifacts/noise from week 1 CBCT to pCT and reconstructing using OS-SART technique. The five different variations correspond to different parameters in the power-law adaptive histogram equalization (PL-AHE) technique. The OAR masks manually segmented on pCT images are perfectly paired to all different psCBCT variations as well.}
    \label{fig:psCBCTVariations}
\end{figure}

To facilitate extraction of CBCT artifacts, week1 baseline CBCTs were deformably registered to their pCTs using a BSpline regularized diffeomorphic image registration~\cite{10.3389/fninf.2013.00039} technique. Since the pCT and baseline week1 CBCT images are acquired on the same day or have few days' difference, the registration artifact mapping accuracy is least susceptible to uncertainties. After image registration, scatter/noise artifacts were extracted from the week1 CBCT using power-law adaptive histogram equalization (PL-AHE)~\cite{PLHAE} that contained the highest to the smoothest frequency components. Different combinations of the parameters of PL-AHE method were used to extract artifacts covering a large range of frequency components. The extracted CBCT artifacts were added to the corresponding pCT and intensities were scaled to $[0, 1]$. Then, 2D x-ray projections were generated from the artifact-induced pCTs using the 3D texture memory linear interpolation of the integrated sinograms. The projections with added gaussian noise were reconstructed using iterative Ordered-Subset Simultaneous Algebraic Reconstruction Technique (OS-SART)~\cite{Wang2004OrderedsubsetSA}, to generate psCBCTs. Finally, we also used geometric data augmentations including, scale ($1.2$)/shear ($8$ degree), and scale ($0.8$)/rotate ($5$ degree). Using this combination of physics-based artifact induction and geometric data augmentations, we can convert a single pCT/week1 CBCT pair into 17 perfectly registered pCT/psCBCT/OAR pairs. Figure~\ref{fig:psCBCTVariations} shows an example of five different psCBCT variations generated from a single deformably registered week 1 CBCT and pCT pair.

\begin{figure}[t!]
\begin{center}
\small
\begin{tabular}{c}
\includegraphics[width=0.8\linewidth]{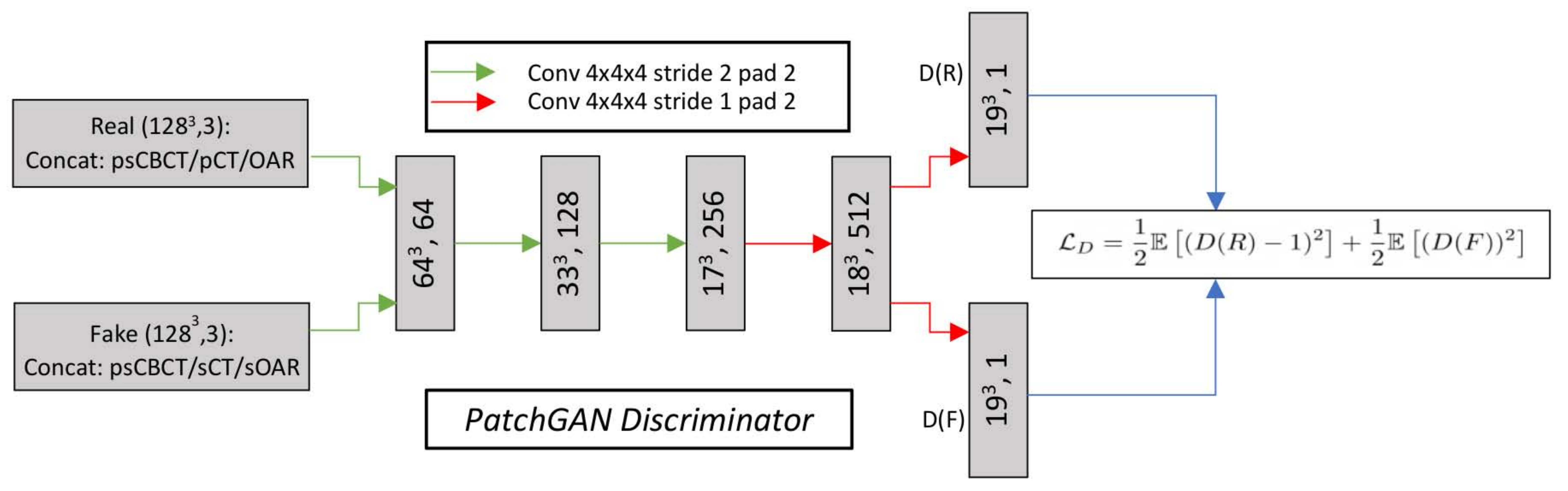}\\
(a) Discriminator pipeline\\
\\
\includegraphics[width=0.8\linewidth]{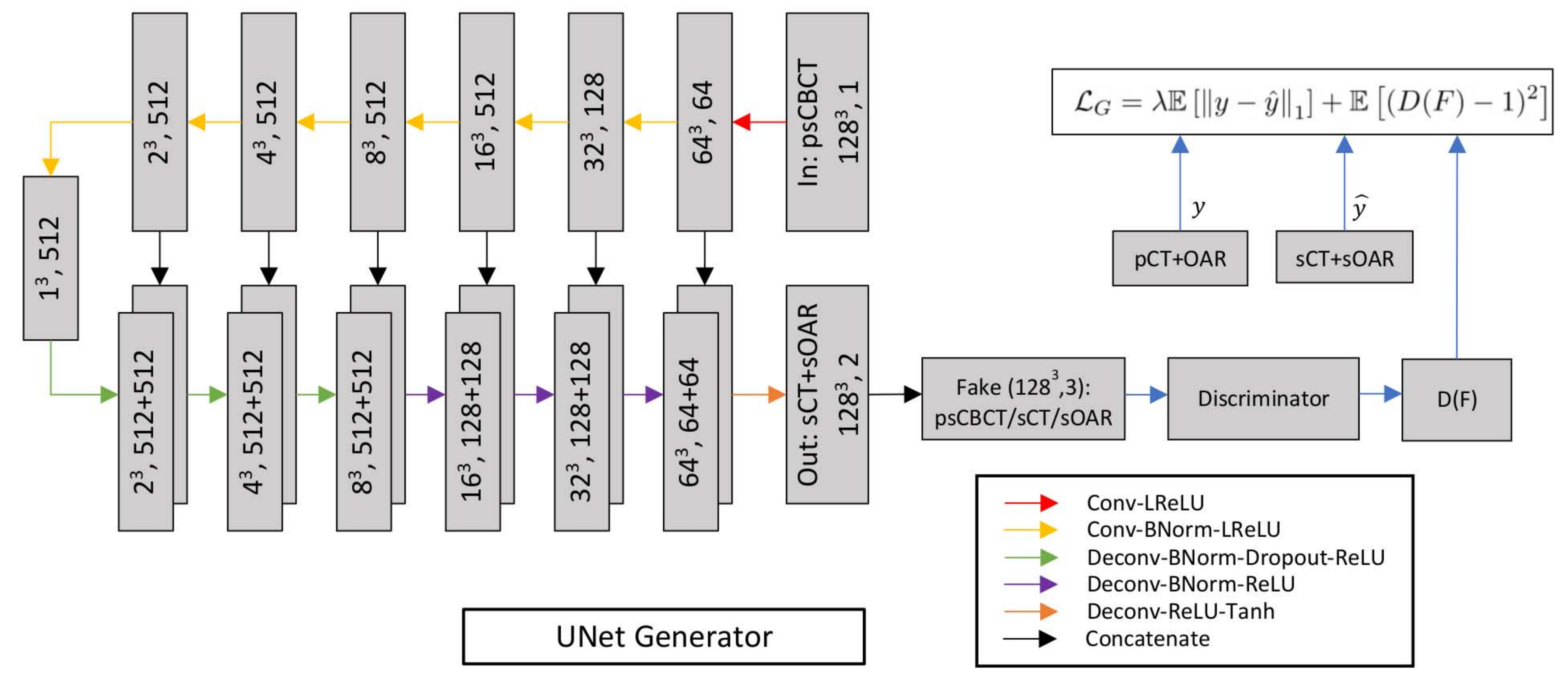}\\
(b) Generator pipeline
\end{tabular}
\caption{(a) $70\times70\times70$ \textit{patch}GAN Discriminator architecture and (b) Unet-like Generator architecture.}
\label{fig:GeneratorDiscriminatorArchs}
\end{center}
\end{figure}

\subsection{Deep Learning Setup}
In this paper, we use the general purpose framework of Image-to-Image translation with Conditional Adversarial Networks (\textit{pix2pix}) for the joint task of CBCT-to-CT translation and OARs segmentation. 

\subsubsection{Conditional Generative Adversarial Networks (cGANs)}
Conditional GANs (cGANs) generate an output conditioned on an input. In cGANs, both the generator and discriminator models are conditioned on ground truth labels or images. More formally, GANs learn a mapping, $G : z \rightarrow y$, where $z$ is a random noise vector and $y$ is the output. Conditional GANs, on the other hand, learn a mapping, $G : \{x, z\} \rightarrow y$, where $x$ is an observed image. So, cGANs learn the mapping conditioned on the input. The generator $G$ is trained to produce real looking outputs which cannot be distinguished, from real images, by an adversarially trained discriminator network $D$. This process can be captured by the following loss function:
\begin{equation}
    \Lagr_{cGAN}(G,D) = \E_{x,y}\left[log(D(x,y)) + \E_{x,z}[log(1-D(x, G(x,z)))\right] \label{eq:cGANLoss}
\end{equation}
where the generator network $G$ tries to minimize this objective against the adversarial network $D$ that tries to maximize it, i.e.
\begin{equation}
    G^{*} = \argmin_G \argmax_D\Lagr_{cGAN}(G,D)
\end{equation}
Along with this cGAN loss, an $L1$ loss is also used on the generator. The generator is tasked to not only fool the discriminator but also produce outputs that are near the input in an $L1$ sense.
\begin{equation}
    \Lagr_{L1}(G) = \E_{x,y,z}[\norm{y - G(x,z)}_1] \label{eq:L1Loss}
\end{equation}
The final combined objective function is given as:
\begin{equation}
G^* = \argmin_G \argmax_D \Lagr_{cGAN}(G, D) + \lambda \Lagr_{L1}(G).    \label{eq:cGANLossFull}
\end{equation}
where $\lambda$ is a tunable hyperparameter to balance the two loss components.
We customize the cGAN framework to translate input 3D psCBCT images to artifact-free 3D synthetic-CT images along with the corresponding OARs segmentation. In our case, the generator takes a 3D psCBCT image ($X: 128^3$) as input and outputs a two channel 3D image ($\hat{Y}: 128^3, 2$) where the channels are composed of sCT and segmented OARs for a given input. Ideally, this output, $\hat{Y}$, would be indistinguishable from the ground truth ($Y: 128^3, 2$), which is the real planning CT and OAR pair concatenated in the channel axis. We want the anatomy in sCT and OARs to correspond to a particular input (psCBCT). This is why we use cGANs which generate new output sample \textit{conditioned} on the input instead of a random noise vector. Since the output is dependent on a particular input, the input psCBCT image is concatenated once with the corresponding ground truth pCT and OAR contours creating the real sample ($R: 128^3, 3$) and once with the generated sCT and OAR contours to create the fake sample ($F: 128^3, 3$). The real and fake samples are input to the discriminator model which predicts how real each input appears ($D(R)$ and $D(F)$). The weights of the generator and discriminator models are then updated using back-propagation based on the loss functions $\Lagr_G$ and $\Lagr_D$ such that the generator produces outputs that match the ground truth inputs leading to the discriminator failing to distinguish between real and fake samples. Instead of the typical loss function for $\Lagr_G$ and $\Lagr_D$, we use a more robust least-squares (LSGAN)~\cite{LSGAN} variation:
\begin{align}
    \Lagr_G &= \E\left[(D(F)-1)^2\right] + \lambda \E\left[\norm{Y - \hat{Y}}_1\right]\\
    \Lagr_D &= \frac{1}{2}\E\left[ (D(R) - 1)^2\right] + \frac{1}{2}\E\left[(D(F))^2\right]
\end{align}
\subsubsection{Generator/Discriminator Architecture}
The generator network follows a common encoder-decoder style architecture which is composed of a series of layers which progressively downsample the input (encoder), until a bottleneck layer, where the process is reversed (decoder). Additionally, Unet-like skip connections are added between corresponding layers of encoder and decoder. This is done to share low-level information between the encoder and decoder counterparts.

The generator (Fig.~\ref{fig:GeneratorDiscriminatorArchs} bottom) uses combinations of Convolution-BatchNorm-ReLU and Convolution-BatchNorm-Dropout-ReLU layers with some exceptions. Batchnorm is not used in the first layer of encoder and all ReLU units in the encoder are leaky with slope of $0.2$ while the decoder uses regular ReLU units. Whenever dropout is present, a dropout rate of $50\%$ is used. All the convolutions in the encoder and deconvolutions (transposed convolutions) in decoder are $4\times4\times4$ 3D spatial filters with a stride of 2 in all 3 directions. The convolutions downsample by $2$ in the encoder and deconvolutions upsample by $2$ in the decoder. Note that, this eliminates the need for separate downsampling or upsampling layers and allows the model to learn up/down sampling functions. The last layer in the decoder maps its input to a two channel output ($128^3, 2$) followed by a $Tanh$ non-linearity.

A discriminator network (Fig.~\ref{fig:GeneratorDiscriminatorArchs} top) termed \textit{Patch}GAN~\cite{DBLP:conf/eccv/LiW16} is used that only penalizes structure at the scale of patches. It is run convolutionally across the input image and tries to classify if each $70\times 70\times 70$ input image patch\footnote{Receptive field calculation shown in appendix.} (3D) is real or fake. All the patch responses are averaged to provide the final output of discriminator $D$. The input to the discriminator can be either the real or fake images. The real input is obtained by concatenating the input psCBCT with CT and OAR images along the channel axis while the fake input is obtained by concatenating the input psCBCT with the translated sCT/sOAR pair output by the generator network. The discriminator's job is to tell apart the real or fake inputs. All ReLUs in the discriminator are leaky, with slope 0.2 and similar to the generator network, batch norm is not applied to the first convolution layer. 

\subsubsection{Input Preprocessing}
The input psCBCT and CT images are in the range $[-1000, 3095]$. In deep learning methods, for faster convergence and numerical stability, input images are generally converted to the range $[0, 1]$. The generator applies a Tanh function (range $[-1, 1]$) to produce its final output. Hence, we apply the following mapping to all our input CT/CBCT images to convert them to $[-1, 1]$ range:
\begin{equation}
    I_{input} = \left( \frac{I + 1000}{4095} \right)*2.0 - 1.0 \label{eq:ImageNormalization}
\end{equation}
We combine the binary RT structure labels for different organs into a single multi-label image with labels: Background = 0, Lungs = 1, Heart = 2, Spinal cord = 3 and Esophagus = 4. We then apply the following transformation to make the multi-label image to have $[-1, 1]$ range:
\begin{equation}
    L_{input} = \left( \frac{L}{4}\right)*2.0 - 1.0 \label{eq:LabelNormalization}
\end{equation}
We deformably register week 1 CBCT (target) to its full FOV pCT (reference) to make sure corresponding OAR structures are well aligned. Then we crop the pCT according to the overlapped FOV. Finally, the images are resized to an isotropic dimension of $128\times 128\times 128$.
The dataset is split into 80 training and 15 testing images. The training set is further divided into 60 training and 20 validation images leading to a 60/20/15(train/valid/test) split. We test various settings on this 60/20 split and after all optimization and thorough experimentation, we train our final models on the 80 images (training + validation), and report the results on the week1 CBCT images in the remaining testing split of 15 images. We use our data augmentation pipeline (physics-based artifact induction + geometric augmentations) to convert the 80 training images into 1360 3D training images. To further improve our results, we later converted the 60 AAPM pCT/OAR cases to 732 psCBCT datasets and added to our training set. Since the streak/scatter artifacts are random, we experimentally chose a CBCT case from our internal dataset and added its different types of extracted artifacts to each AAPM CT cases. The artifact images were first deformably registered to each case to align them into the same coordinate system (same origin, voxel size, dimension, etc). Then we simply perform a pixelwise addition of the CTs and artifact-only images in the overlapped area (numeric conversion [casting] is done to make sure they have same pixel type). Then, the resulting artifact-induced CT intensities are rescaled to [0,1]. Eventually x-ray projections were generated from the intensity rescaled artifact-induced CTs and reconstructed using OS-SART technique to produce synthetic-CBCTs for the entire AAPM dataset.

\begin{figure}[t!]
\begin{center}
\small
\begin{tabular}{c}
\includegraphics[width=0.9\linewidth]{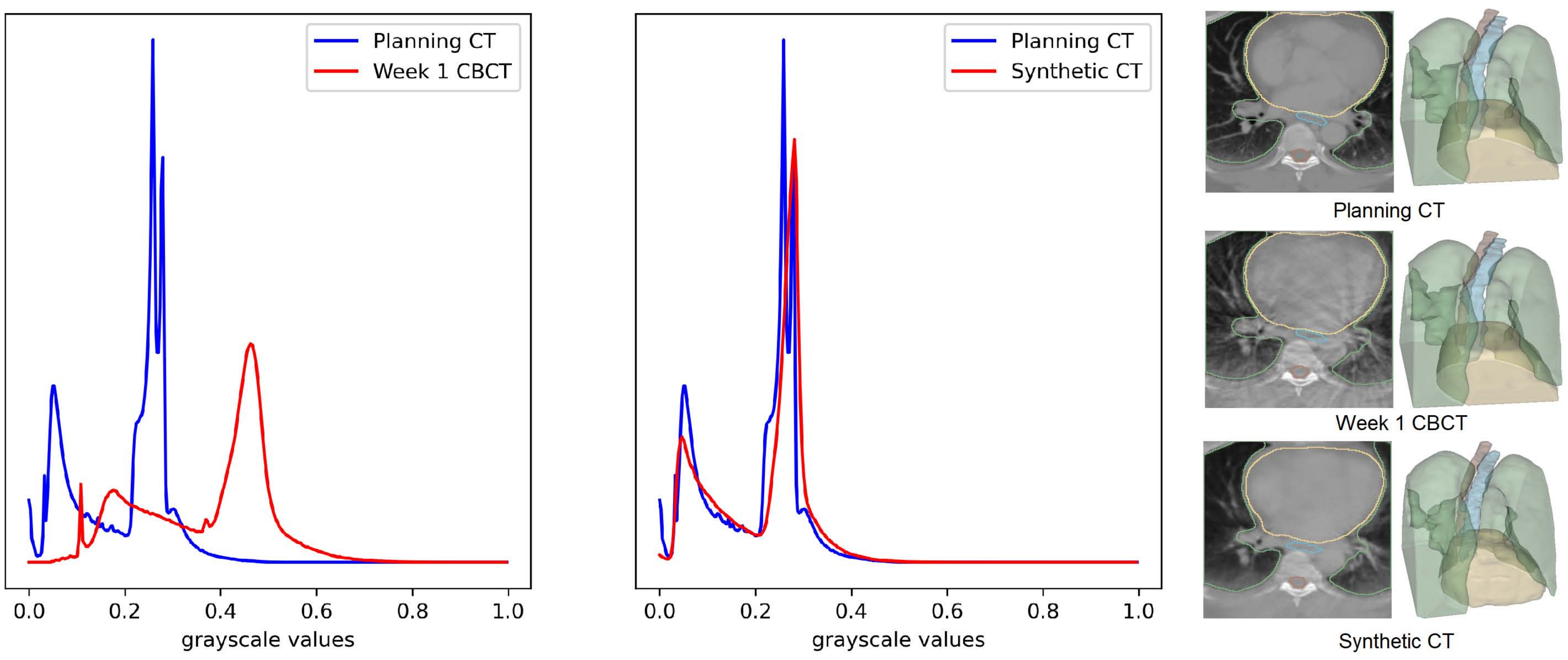}\\
\end{tabular}
\caption{Planning CT, week 1 CBCT, the generated synthetic CTs and the corresponding OARs. The histogram of the synthetic CT is clearly aligned with planning CT compared to Week 1 CBCT.}
\label{fig:psCBCT_histograms}
\end{center}
\end{figure}

\begin{figure}[t!]
\begin{center}
\small
\begin{tabular}{c}
\includegraphics[width=\linewidth]{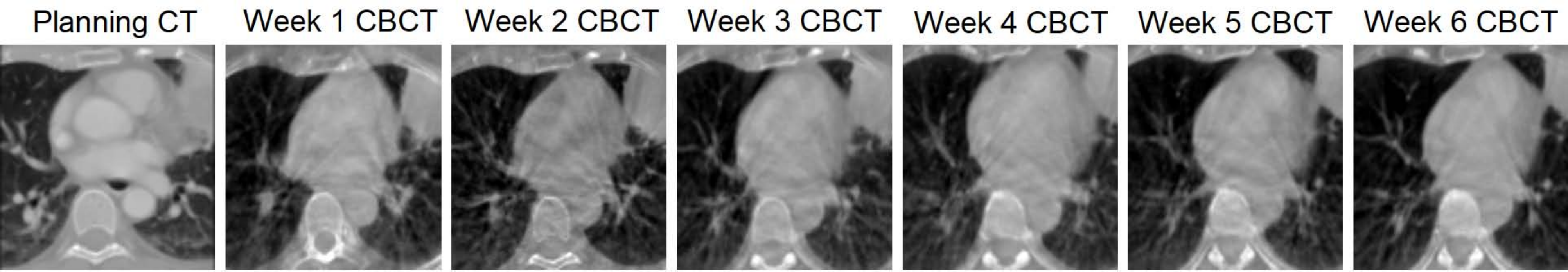}\\
\includegraphics[width=\linewidth]{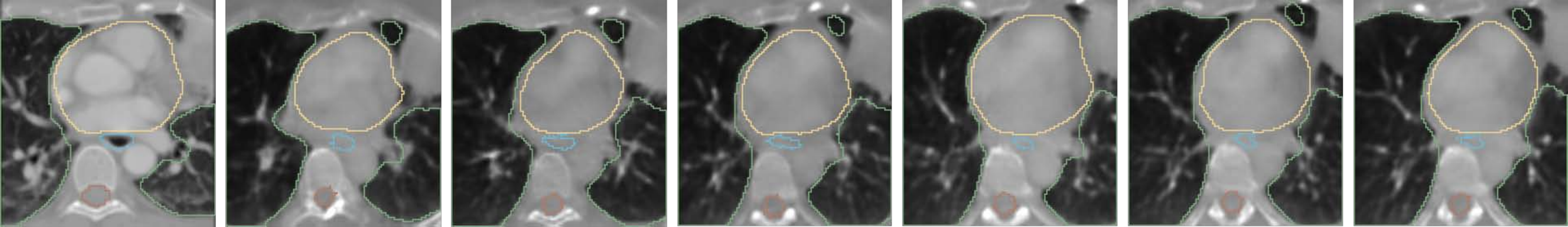}\\
\includegraphics[width=\linewidth]{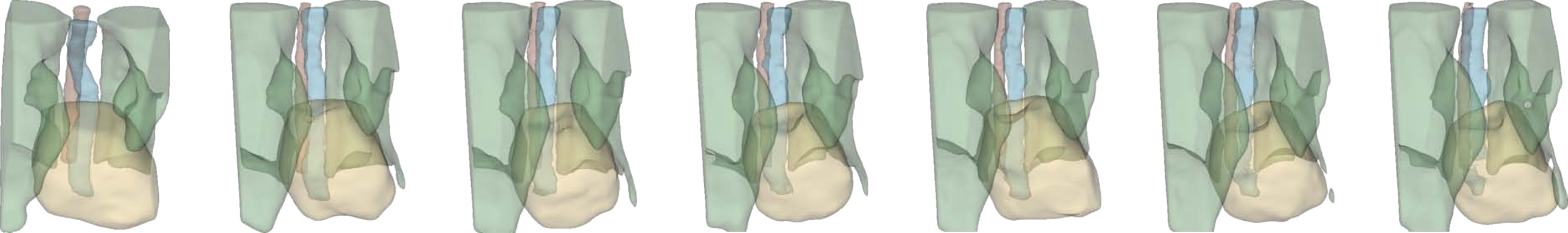}
\end{tabular}
\caption{Synthetic CTs (middle) and OAR segmentations (bottom) generated for real weekly CBCTs (top).}
\label{fig:fig:Week1-Week6-Result}
\end{center}
\end{figure}

\subsubsection{Settings}
In our experiments, used to report the final results, we use Stochastic Gradient Descent (SGD), with a batch size of 1, and Adam optimizer~\cite{DBLP:journals/corr/KingmaB14} with an initial learning rate of $0.0002$, and momentum parameters $\beta_1 = 0.5$, $\beta_2 = 0.999$. We train the network for total of 100 epochs. We use a constant learning rate of $0.0002$ for the first 50 epochs and then let the learning rate linearly decay to 0 for the final 50 epochs. We use $\lambda = 100$, as a balancing factor between cGAN and L1 losses.

A common problem in using GAN based methods, especially for image segmentation (discrete label generation) tasks, is that the discriminator can differentiate between real and fake outputs very easily. This leads to severe instability in training the networks. We use a combination of several strategies~\cite{arjovsky2017wasserstein, salimans2016improved} to overcome this issue. We use label smoothing whereby instead of using 0/1 to represent fake/real samples, we use random numbers between 0 to 0.3 for fake labels and between 0.7 to 1.2 for real labels. We also use the strategy of confusing the discriminator by randomly exchanging the real and fake images before passing to the discriminator. During the optimization phase, we exchange real and fake images with a probability of $10\%$. Finally, we inject noise to the input of the discriminator by adding zero mean gaussian noise to the real/fake images before feeding to the discriminator. This noise is annealed linearly over the course of training, starting with a variance of 0.2 it goes linearly down to 0 at the final epoch of training. This leads to stabilization of the training process with better gradient back-propagation.

\subsubsection{Model implementation and training}
We created the implementations of the multi-task 3D Unet-like generator, \textit{patch}GAN discriminator, $L1$ and cGAN loss functions, GAN stabilizing techniques and all other related training/testing scripts in pytorch and we conducted all our experiments on a Nvidia RTX 2080 Ti GPU with 11 GB VRAM.

%% file: experimentalresults.tex
\subsection{Qualitative results}
Figure \ref{fig:psCBCT_histograms} shows planning CT, registered week 1 CBCT and our synthetic CT output along with the corresponding OARs for one of the test images. The week 1 CBCT translation to synthetic CT brings the intensity-value histogram closer to planning CT images, showing that the trained model has learnt to remove the CBCT scatter/noise artifacts. We can however also note that, the generated sCT image has somewhat less contrast than the real CT image. We believe this is due to smoothing effect of the L1 loss. We can also see the slight smoothing effect of the translation process in the planning CT/synthetic CT histogram: real CT has two distinct peaks which are somewhat averaged into a single peak in the translated CT. Finally, we show an example \ref{fig:fig:Week1-Week6-Result} of synthetic CTs and OARs generated for real week 1 to week 6 CBCT images.

\subsection{Quantitative results}
\bgroup
\def\arraystretch{1.5}
\begin{table}[ht!]
\small
\begin{center}
\caption{Evaluation criteria for comparing quality of generated sCT with the real CT compared to the input real week 1 CBCT. Mean absolute error and RMSE are given in CT Hounsfield Units (HU) with respect to pCT values. msk and aapm refers to our internal data and the physics-augmented AAPM challenge data. eso1 refers to experiments where esophagus was labeled 1 and eso4 to the experiment where the labels were flipped and esophagus was labeled 4. The stabilized version with GAN stabilization techniques added to the optimization phase produces better results compared to the unstabilized version.
\label{tab:cbct2ctcriteria}
\vspace*{2ex}
}
\begin{tabular}{|l|c|c|c|c|}
\hline
\multirow{2}{*}{Modality/Settings} & MSSIM                               & \multicolumn{1}{l|}{MAE (HU)}       & \multicolumn{1}{l|}{PSNR (dB)}      & \multicolumn{1}{l|}{RMSE (HU)}      \\ \cline{2-5} 
                                   & \multicolumn{1}{l|}{mean $\pm$ std} & \multicolumn{1}{l|}{mean $\pm$ std} & \multicolumn{1}{l|}{mean $\pm$ std} & \multicolumn{1}{l|}{mean $\pm$ std} \\ \hline
w1CBCT                             & 0.73 $\pm$ 0.07                     & 162.77 $\pm$ 53.91                  & 22.24 $\pm$ 2.40                    & 328.18 $\pm$ 84.65                  \\
sCT (msk+aapm+stabilized+eso4)     & 0.92 $\pm$ 0.01                     & 29.31 $\pm$ 12.64                   & 34.69 $\pm$ 2.41                    & 78.62 $\pm$ 24.22                   \\
sCT (msk+aapm+unstabilized+eso4)   & 0.88 $\pm$ 0.02                     & 39.19 $\pm$ 19.95                   & 32.89 $\pm$ 3.08                    & 99.15 $\pm$ 38.32                   \\
sCT (msk+stabilized+eso4)          & 0.91 $\pm$ 0.02                     & 40.06 $\pm$ 22.72                   & 32.93 $\pm$ 3.39                    & 100.03 $\pm$ 41.57                  \\
sCT (msk+aapm+stabilized+eso1)     & 0.90 $\pm$ 0.02                     & 43.57 $\pm$ 22.72                   & 32.80 $\pm$ 3.79                    & 102.84 $\pm$ 42.93                  \\ \hline
\end{tabular}
\end{center}
\end{table}
\egroup
We used mean structural similarity index measure (MSSIM), mean absolute error (MAE), root mean square error (RMSE) and peak signal-to-noise ratio (PSNR) criteria to measure the similarity of the generated sCT (from real week 1 CBCT) with pCT images; MAE and RMSE is reported in CT Hounsfield Units (HU). The metrics are reported in Table~\ref{tab:cbct2ctcriteria}. As shown, the addition of publicly available AAPM data augmented with our physics-based technique to our training improved the results for both translation and segmentation tasks. Without GAN stabilizing techniques poor results are obtained for both the tasks.

\bgroup
\def\arraystretch{1.5}
\begin{table}[htb!]
\small
\caption{Evaluation criteria for comparing quality of generated segmentation masks for organs-at-risk compared to manual segmentations for real week1 CBCT input images. Setting anatomy labels for esophagus, spinal cord, heart and lungs as 4, 3, 2, 1 respectively (instead of 1, 2, 3, 4 respectively) helps to substantially improve the segmentation results for esophagus and spinal cord without any noticeable degradation in heart and lungs results. The stabilized version has GAN stabilization techniques added in the optimization phase and produces better results compared to the unstabilized version.
\label{tab:segmentationcriteria}
\vspace*{2ex}
}
\begin{center}
\begin{tabular}{|l|l|c|c|c|}
\hline
\multirow{2}{*}{Settings}                & \multirow{2}{*}{Anatomy} & DICE                                & \multicolumn{1}{l|}{MSD (mm)}       & \multicolumn{1}{l|}{HD95 (mm)}      \\ \cline{3-5} 
                                         &                          & \multicolumn{1}{l|}{mean $\pm$ std} & \multicolumn{1}{l|}{mean $\pm$ std} & \multicolumn{1}{l|}{mean $\pm$ std} \\ \hline
\multirow{4}{*}{msk+aapm+stabilized+eso4} & Lungs                    & 0.96 $\pm$ 0.01                     & 0.96 $\pm$ 0.20                     & 3.44 $\pm$ 0.93                     \\
                                         & Heart                    & 0.88 $\pm$ 0.08                     & 2.28 $\pm$ 0.94                     & 8.07 $\pm$ 4.80                     \\
                                         & Spinal Cord              & 0.83 $\pm$ 0.03                     & 1.12 $\pm$ 0.32                     & 3.45 $\pm$ 3.39                     \\
                                         & Esophagus                & 0.66 $\pm$ 0.06                     & 2.22 $\pm$ 0.40                     & 6.50 $\pm$ 1.91                     \\ \hline
\multirow{4}{*}{msk+aapm+unstabilized+eso4} & Lungs                    & 0.95 $\pm$ 0.01                   & 1.38 $\pm$ 0.55                     & 6.34 $\pm$ 4.83                     \\
                                         & Heart                    & 0.81 $\pm$ 0.09                     & 4.24 $\pm$ 1.32                     & 19.08 $\pm$ 7.19                     \\
                                         & Spinal Cord              & 0.80 $\pm$ 0.03                     & 1.25 $\pm$ 0.39                     & 3.81 $\pm$ 3.97                     \\
                                         & Esophagus                & 0.58 $\pm$ 0.11                     & 4.01 $\pm$ 2.96                     & 17.83 $\pm$ 20.73                    \\ \hline
\multirow{4}{*}{msk+stabilized+eso4}      & Lungs                    & 0.95 $\pm$ 0.01                     & 1.01 $\pm$ 0.21                     & 3.71 $\pm$ 1.08                     \\
                                         & Heart                    & 0.88 $\pm$ 0.10                     & 2.13 $\pm$ 0.82                     & 7.38 $\pm$ 3.27                     \\
                                         & Spinal Cord              & 0.82 $\pm$ 0.03                     & 1.16 $\pm$ 0.36                     & 3.61 $\pm$ 3.68                     \\
                                         & Esophagus                & 0.63 $\pm$ 0.06                     & 2.51 $\pm$ 0.45                     & 8.27 $\pm$ 4.08                     \\ \hline
\multirow{4}{*}{msk+aapm+stabilized+eso1} & Lungs                    & 0.96 $\pm$ 0.01                     & 1.03 $\pm$ 0.24                     & 3.73 $\pm$ 1.08                     \\
                                         & Heart                    & 0.87 $\pm$ 0.08                     & 2.31 $\pm$ 0.78                     & 7.43 $\pm$ 2.21                     \\
                                         & Spinal Cord              & 0.78 $\pm$ 0.04                     & 1.32 $\pm$ 0.36                     & 3.81 $\pm$ 2.87                     \\
                                         & Esophagus                & 0.56 $\pm$ 0.12                     & 3.61 $\pm$ 2.26                     & 15.22 $\pm$ 15.76                     \\ \hline
\end{tabular}
\end{center}
\end{table}
\egroup
To measure the accuracy of the generated organs-at-risk segmentations, we computed DICE coefficient, mean surface distance (MSD) and 95\% Hausdorff distance (HD95), shown in Table~\ref{tab:segmentationcriteria}. The heart, lungs, and spinal cord segmentation results are close to the state-of-the-art~\cite{https://doi.org/10.1002/mp.13141}. Esophagus is the most challenging of the OAR structures given the thin tubular structure. Addition of the physics-augmented AAPM data helped improve segmentation of esophagus with the DICE increasing from 0.63 to 0.66. A key element in improving the segmentation results, particularly for spinal cord and esophagus turned out to be the numerical labels used for different anatomies. In one experiment, we used the labels 1,2,3,4 (\textbf{eso1}) for esophagus, spinal cord, heart and lungs respectively, and in another we flipped the order to 4,3,2,1 (\textbf{eso4}). The order with higher label values for esophagus and spinal cord substantially increased their respective segmentation scores. Using \textbf{eso4} label order resulted in average DICE of 0.66 and 0.84 for esophagus and spinal cord respectively, up from 0.56 and 0.58 respectively for the \textbf{eso1} label order. This increase primarily comes from the fact that $L1$ loss is used in the generator. In general, a mismatch in the actual and generated values for the anatomy with higher label value will result in higher $L1$ loss value and in turn greater gradient back-propagation. Lungs and heart structures occupy much larger volume compared to esophagus and spinal cord which are thinner tubular structures. Hence, there is natural class imbalance in the $L1$ loss function. Ideally, a loss function which produces identical loss magnitude regardless of the label of the structure would be more desirable. In the context of L1 loss in our framework this could be achieved by using one hot encoding of the structures instead of assigning numerical labels to the different structures. With one hot encoding reweighting of the loss magnitude could be possible using the inverse of real volume to deal with the class imbalance issue.The one-hot encoding scheme failed to produce results owing to the inherent unstability in training GAN models. In our case, it becomes trivially easy for the discriminator to distinguish the real and fake structures especially the smaller ones (esophagus and spinal cord). By combining the structures into a single multilabel image combined with the training stabilization techniques we were able to overcome this difficulty. With this strategy, once the training had stabilized we dealt with class imbalance issue of relative loss weighting between different structures by modifying the structure labeling order. In future work, we will focus our efforts on developing a loss function which is naturally suited to both the tasks of general image-to-image translation (CBCT-to-CT) and segmentations while simultaneously incorporating features to help mitigate the class imbalance issue of segmentation.

%% file: discussion.tex
\begin{figure}[htb!]
    \centering
    \includegraphics[width=\linewidth]{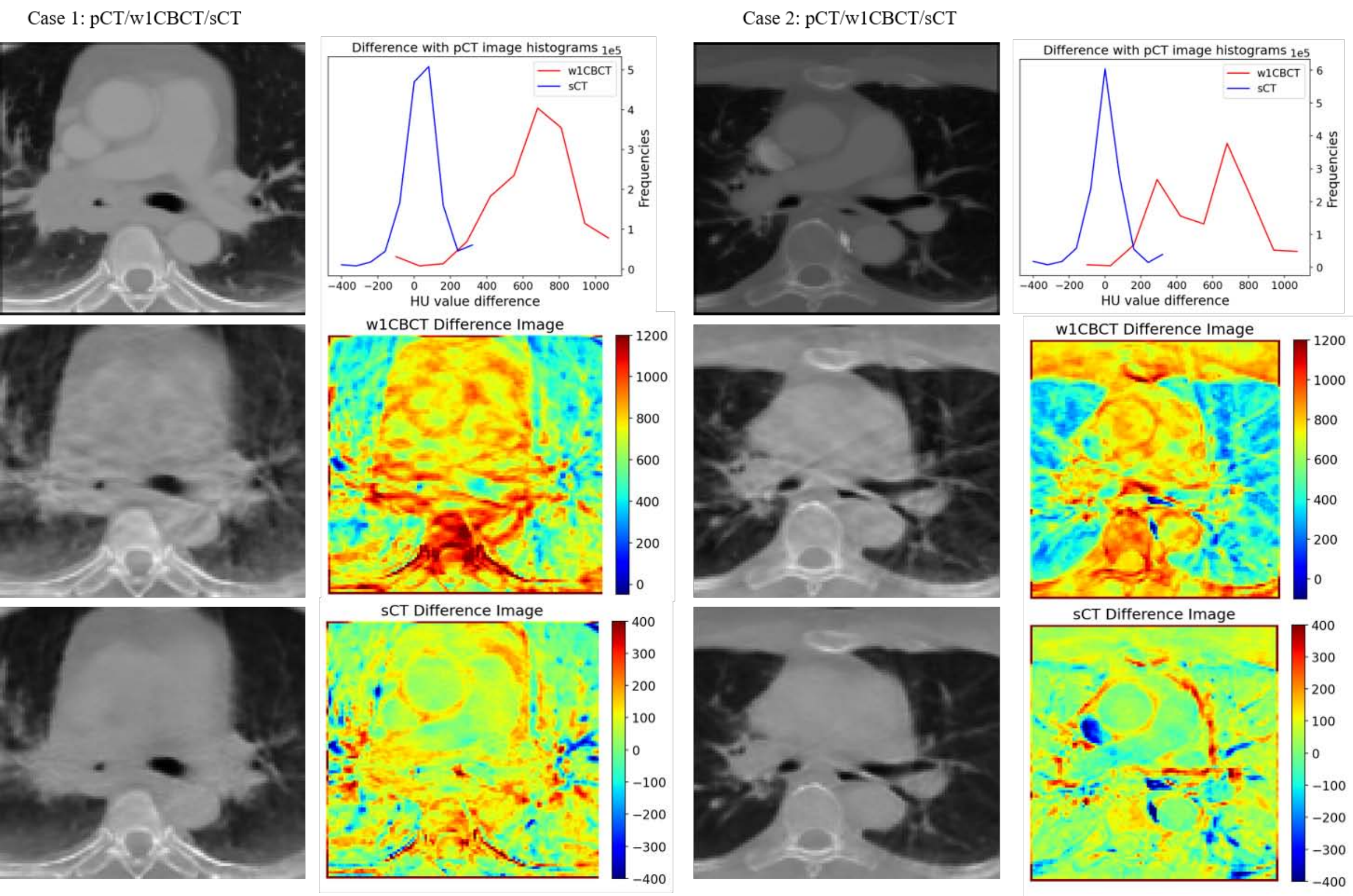}%
    \caption{Planning CT, w1CBCT and sCT slices for two patient cases along with residual images between w1CBCT/pCT (MAE 216.19 HU) and sCT/pCT (MAE 31.47 HU). Histograms of residual images show that the sCT image residual errors are concentrated around 0 HU largely contained within $\pm$ 100 HU whereas w1CBCT images show extremely large differences compared to pCT.}
    \label{fig:DifferenceImages}
\end{figure}

Weekly/daily CBCT images are currently used for patient setup only during image-guided radiotherapy. In this paper, we presented an approach to help quantify CBCT images by converting them to high-quality synthetic CT images while simultaneously generating segmentation masks for organs-at-risk including lungs, heart, spinal cord and esophagus. 
While our approach produces high quality results, nevertheless there are some drawbacks worth mentioning. One of issues relates to the use of $L1$ loss in the generator network which causes blurred results. This results in lower contrast in the generated sCT compared to pCT. A large proportion of the residual errors in the sCT images are concentrated on or near the edges of different anatomical regions, as shown in figure~\ref{fig:DifferenceImages}. We tried to fix this issue by using combination of linear or nearest neighbor upsampling and convolution instead of transposed convolutions in the generator decoder. This modification indeed increased the contrast in the sCT images, however, it led to a significant decrease in the segmentation results of the smaller anatomies, namely, esophagus and spinal cord. In the future, we will look for more principled ways to overcome this problem including alternative loss functions.


Finally, all our results, including CBCT to CT translation, lungs, heart and spinal cord segmentation, except esophagus segmentation, are close to the state-of-the-art results for these individual tasks using different approaches and models. For further improvements, we will focus our efforts to develop ways to improve the esophagus segmentation to state-of-the-art levels as well. In that direction, our previous work~\cite{Sadegh-psCBCT} used a modified 3D Unet architecture with the physics-based data augmentation approach to achieve state-of-the-art segmentation results on esophagus. To improve our segmentation results, we will change our current generator architecture in line with the modified 3D Unet. Furthermore, with the addition of the publicly available AAPM training data, we saw improved results across tasks with our model i.e. both decreased MAE on sCT images and increased DICE scores on OAR segmentation. In the future, we plan to use physics-augmented 422 non-small cell lung cancer patients data from publicly available The Cancer Imaging Archive (TCIA)~\cite{TCIV} to help improve results across tasks. Another major enhancement in future work will be to add an additional task of radiation dose inference along with the current tasks of simultaneous sCT generation and OAR segmentation.

%% file: conclusions.tex
In this paper we presented an approach to help quantify noisy artifact-ridden CBCT images by converting them to high quality synthetic CT images while simultaneously segmenting organs-at-risk including lungs, heart, spinal cord and esophagus. This approach can allow clinicians to use weekly/daily CBCT images for adapting radiotherapy mid-course, given the anatomical changes, to potentially improve patient outcomes.

%% file: appendix.tex
We use the following definitions and formulae for the evaluation criteria of synthetic CT generation task and organs-at-risk segmentation:

\begin{enumerate}
    \item MSSIM: It is an objective measure to characterize the perceived quality of an image compared to a reference image. It considers the idea that spatially close pixels have strong inter-dependencies. The following expression is used in the computation of MSSIM:
    \begin{equation}
        SSIM(x,y) = \frac{\left(2\mu_x \mu_y + C_1\right)\left(2\sigma_{xy} + C_2\right)}{\left(\mu_x^2 + \mu_y^2 + C_1\right)\left(\sigma_x^2 + \sigma_y^2 + C_2\right)}
    \end{equation}
    where $x,y$ are two images, $\mu$ is the mean of an image, $\sigma$ is the standard deviation, $\sigma_{xy}$ is covariance between two images, $C_1,C_2$ are constants with $C_1 = 0.01^2$ and $C_2 = 0.03^2$. In our implementation we use a sliding window approach to compute local SSIM and take average to report mean SSIM (MSSIM).
    
    \item MAE: Mean absolute error computes the average of absolute errors of all corresponding pixels in the two images. A lower MAE indicates improved quality and is calculated using the following expression:
    \begin{equation}
        MAE(I_1, I_2) = \frac{1}{N} \sum_{i = 1}^N | I_1(i) - I_2(i)|
    \end{equation}
    where $I_1$ and $I_2$ are two images with $N$ number of total pixels.
    
    \item RMSE: It is commonly used to measure the difference between observed values and those predicted by a model. A lower RMSE indicates better match between two images and is calculated using the following expression:
   
    \item PSNR: PSNR is typically used to measure the quality of results produced by noise reduction methods. Increasing PSNR indicates better quality match between reference and target images. It is calculated as follows:
    \begin{equation}
        PSNR(I_1, I_2) = 20\log_{10}MAX\left( I_r\right) - 10 \log_{10}\left(MSE\left(I_1,I_2\right)\right)
    \end{equation}
    where $I_r$ is one of $I_1$ or $I_2$ to be used as reference image. In our case we will use the pCT as reference and compare the generated synthetic CT against it.
    
    \item DICE: The DICE coefficient is routinely used in medical imaging applications to measure the quality of segmentations. It measures the overlap between a reference and algorithmically generated binary segmentation using the following formula:
    \begin{equation}
        DICE = \frac{2|X\cap Y|}{|X| + |Y|}
    \end{equation}
    
    \item MSD: Mean surface distance (MSD) between a surface $S$ from an automatic method and a reference surface $S_{ref}$ is defined as:
    \begin{equation}
        MSD = \frac{\bar{d}\left(S, S_{ref}\right) + \bar{d}\left(S_{ref}, S\right)}{2}
    \end{equation}
    where $\bar{d}\left(S, S_{ref}\right)$ is the average of distances from \textit{every} voxel in $S$ to its closest voxel in $S_{ref}$ and $\bar{d}\left(S_{ref}, S\right)$ is calculated similarly but is directed from $S_{ref}$ to $S$.
    
    \item HD95: The Hausdorff distance (HD) between a surface $S$ from an automatic method and a reference surface $S_{ref}$ is defined as the maximum of distances from \textit{every} voxel in $S$ to its closest voxel in $S_{ref}$. The 95 percentile Hausdorff distance (HD95) is the voxel in $S$ with distance to its closest voxel in $S_{ref}$ greater or equal to $95\%$ of other voxels in $S$. It is a directed measure and the undirected version is defined as the average of directed HD95 from $S$ to $S_{ref}$ and vice versa. HD95 distance is more robust to outliers than simple HD.
\end{enumerate}

\paragraph{Receptive field calculation of PatchGAN discriminator}The receptive field is calculated between the output of a layer and its input. For example, we could ask how many features in the layer right before the final output layer of the discriminator affect the output at any of the 1x1x1 locations in the output layer? The answer is given by the relationship: $R_{l-1} = K_l + S_l*(R_l - 1)$ where $R_l$ is the receptive field, $S_l$ is the stride and $K_l$ is the kernel size at layer $l$ (https://distill.pub/2019/computing-receptive-fields/). For the discriminator architecture shown in Figure 3(a), the layers may be numbered from $l = 0$ at the input to $l = 5$ at the output. We may then apply the above relationship recursively to map the receptive field of a 1x1x1 location in the output to the input image. The kernel size and strides are provided in the figure for each layer. If the kernel sizes and strides are different in different dimensions then the relation can be used separately for each dimension. In our case all kernels and strides are symmetric in all three dimensions. As an example, for the output layer, $S_5 = 1$, $K_5 = 4$ and $R_5 = 1$ (we are looking at 1x1x1 location in the output). The calculations for all the layers are as follows:
\begin{enumerate}
    \item $R_4 = 4 + 1*(1 - 1) = 4$
    \item $R_3 = 4 + 1*(4 - 1) = 7$
    \item $R_2 = 4 + 2*(7 - 1) = 16$
    \item $R_1 = 4 + 2*(16 - 1) = 34$
    \item $R_0 = 4 + 2*(34 - 1) = 70$
\end{enumerate}
Hence, a 1x1x1 location in the output of the discriminator is dependent on a 70x70x70 patch of the input image (receptive field).